\documentclass[12pt]{article}
\usepackage{amssym}

\font\elevenof=msbm10 at 11pt
\font\bdi=cmmib10 at 12 pt

\def\Z{\mbox{$\Bbb Z$}}
\def\I{\mbox{$\Bbb I$}}
\def\Qp{Q^{\dagger}}
\def\bp{b^{\dagger}}
\def\ap{a^{\dagger}}
\def\cp{c^{\dagger}}
\def\bt{\tilde{b}}
\def\btp{\tilde{b}^{\dagger}}
\def\Qt{\tilde{Q}}
\def\Qtp{\tilde{Q}^{\dagger}}
\def\Ht{\tilde{\cal H}}
\def\Qb{\bar{Q}}
\def\Qbp{\bar{Q}^{\dagger}}
\def\Hb{\bar{H}}
\def\Eb{\bar{\cal E}}
\def\case#1#2{{\textstyle{#1\over #2}}}
\def\diag{\mathop{\rm diag}\nolimits}
\def\mod{\mathop{\rm mod}\nolimits}
\def\AG{${\cal A}(G(N))$}
\def \Athree{${\cal A}^{(3)}(G(N))$}
\def\bi#1{\hbox{\bdi #1\/}}

\newcommand{\xb}{\bi{x}}

\title{
PSEUDOSUPERSYMMETRIC QUANTUM MECHANICS: GENERAL CASE,
ORTHOSUPERSYMMETRIES, REDUCIBILITY, AND BOSONIZATION}
\author{C. QUESNE\thanks{Directeur de recherches FNRS} \ and N. VANSTEENKISTE \\
{\small \sl Physique Nucl\'eaire Th\'eorique et Physique Math\'ematique,}\\ {\small \sl
Universit\'e Libre de Bruxelles, Campus de la Plaine CP229,} \\ {\small \sl  Boulevard~du
Triomphe, B-1050 Brussels, Belgium} \\
{\small \sl E-mail: cquesne@ulb.ac.be}}
\date{ }
\begin{document}
\baselineskip=22pt plus 1pt minus 1pt
\maketitle

\begin{abstract} 
Pseudosupersymmetric quantum mechanics (PsSSQM), based upon the use of
pseudofermions, was introduced in the context of a new Kemmer equation describing
charged vector mesons interacting with an external constant magnetic field. Here we
construct the complete explicit solution for its realization in terms of two superpotentials,
both equal or unequal. We prove that any orthosupersymmetric quantum mechanical
system has a pseudosupersymmetry and give conditions under which a
pseudosupersymmetric one may be described by orthosupersymmetries of order two. We
propose two new matrix realizations of PsSSQM in terms of the generators of a generalized
deformed oscillator algebra (GDOA) and relate them to the cases of equal or unequal
superpotentials, respectively. We demonstrate that these matrix realizations are fully
reducible and that their irreducible components provide two distinct sets of bosonized
operators realizing PsSSQM and corresponding to nonlinear spectra. We relate such results
to some previous ones obtained for a GDOA connected with a $C_3$-extended oscillator
algebra (where $C_3 = \mbox{\elevenof Z}_3$) in the case of linear spectra.
\end{abstract}

\noindent
Running head: Pseudosupersymmetric Quantum Mechanics

\noindent
PACS: 03.65.Fd, 11.30.Pb

\noindent
Keywords: quantum mechanics, supersymmetry, generalized deformed oscillator
\newpage
%
%
\section{Introduction}

During the past two decades, supersymmetry, based upon a symmetry between bosons
and fermions~\cite{witten81}, has found a lot of applications in quantum
mechanics~\cite{cooper}. The success of this new field, supersymmetric quantum
mechanics (SSQM), has triggered the search for generalizations by extending the
symmetry to some exotic statistics. Replacing fermions by parafermions~\cite{green},
pseudofermions~\cite{beckers95a}, or orthofermions~\cite{mishra}, for instance, has led
to parasupersymmetric (PSSQM)~\cite{rubakov, beckers90}, pseudosupersymmetric
(PsSSQM)~\cite{beckers95a, beckers95b}, or orthosupersymmetric quantum mechanics
(OSSQM)~\cite{khare}, respectively. Substituting a $\Z_k$ grading to the $\Z_2$ one
characteristic of SSQM has also given rise to fractional supersymmetric quantum
mechanics (FSSQM)~\cite{durand}. More recently, from a somewhat different viewpoint,
extending Witten's index~\cite{witten82} to more general topological invariants has
resulted in the concept of topological symmetries (TS)~\cite{mosta}.\par
%
%
In the present paper, we will come back to one of the generalizations of SSQM, namely
PsSSQM, which has been introduced in the context of a new relativistic Kemmer equation
describing charged vector mesons interacting with an external constant magnetic
field~\cite{beckers95a}. This equation, which has solved for the first time the
longstanding problems of reality of energy eigenvalues and causality of propagation, leads
in the nonrelativistic limit to a pseudosupersymmetric oscillator Hamiltonian, which can be
realized in terms of boson-like operators and pseudofermionic ones, where the latter are
intermediate between fermionic and order-two parafermionic operators.\par
%
%
Later on, PsSSQM has been reformulated in terms of two superpotentials $W_1$ and
$W_2$, but only a special case corresponding to the choice $W_1 = W_2 = W$ has
actually been studied in detail~\cite{beckers95b}. One of the purposes of the present
paper is to derive explicit forms of the pseudosupersymmetric Hamiltonian in the general
case, including both the choices $W_1 = W_2$ and $W_1 \ne W_2$.\par
%
%
Another aim is to reconsider the connections between PsSSQM and OSSQM, partly
analyzed in Ref.~\cite{beckers95b}. Here they will be thoroughly discussed by using a
slightly different approach, thereby emphasizing some similarities with the links between
PSSQM or FSSQM and OSSQM, which have been recently established~\cite{agha}.\par
%
%
Still another purpose is to examine the realizations of PsSSQM in terms of superpotentials
at the light of a recent work, wherein we have provided a bosonization (i.e., a realization
in terms of only boson-like operators without fermion-like ones) of several variants of
SSQM~\cite{cq00}, generalizing a well-known result for SSQM~\cite{plyu} in terms of the
Calogero-Vasiliev algebra~\cite{vasiliev}. For the SSQM variants, the algebras used are
some generalized deformed oscillator algebras (GDOAs) (see Ref.~\cite{katriel} and
references quoted therein) related to $C_{\lambda}$-extended oscillator
ones~\cite{cq00, cq98}, where $C_{\lambda} = \Z_{\lambda}$ denotes the cyclic
group of order $\lambda$ ($\lambda \in \{3, 4, 5, \ldots\}$). Such GDOAs reduce to
the Calogero-Vasiliev algebra for $\lambda = 2$. In the case of PsSSQM, we plan to
establish here a correspondence between the two matrix realizations obtained for $W_1
= W_2$ and $W_1 \ne W_2$, respectively, and the two different types of bosonization
obtained in Ref.~\cite{cq00}.\par
%
%
A fourth purpose of the present work is to go further than the oscillator spectra
considered in Ref.~\cite{cq00} by providing two new matrix realizations of PsSSQM in
terms of GDOA generators in the case of general nonlinear spectra. Such matrix
realizations will prove fully reducible and will lead to two different types of PsSSQM
bosonization valid for nonlinear spectra.\par
%
%
This paper is organized as follows. In Sec.~2, we review the physical motivation for the
introduction of PsSSQM. In Sec.~3, we obtain two different matrix realizations in terms
of superpotentials. In Sec.~4, we provide an entirely new analysis of the connections
between PsSSQM and OSSQM. We introduce two matrix realizations of PsSSQM in terms of
GDOA generators in Sec.~5 and use them in Sec.~6 to establish the reducibility and
bosonization of PsSSQM. In Sec.~7, the results of Secs.~5 and 6 are specialized to GDOAs
associated with $C_3$-extended oscillator algebras. Finally, Sec.~8 contains a summary
of the main results.\par
%
%
\section{Physical Motivation for the Introduction of Pseudosupersymmetric Quantum
Mechanics}

Until recently, the problem of the interaction of relativistic vector mesons with an
external electromagnetic field has been plagued with two main difficulties: the existence
of complex energy eigenvalues and the violation of the causality principle (see \cite{vija}
and references quoted therein). To eliminate such drawbacks, it has been proposed to add
a new term characterized by some real parameter $\lambda$ to the Kemmer equation
describing the phenomenon~\cite{beckers95a}.\par
%
%
In the simplest context of a constant magnetic field, the modified Kemmer equation,
when reduced to its Sakata-Taketani form, gives rise to a six-component Klein-Gordon
type equation
\begin{equation}
  P_0^2 \chi(\xb) = \left(\vec{\Pi}^2 + 1 - 2eB \Sigma_3 + \lambda eB\right) \chi(\xb),
  \label{eq:KG}
\end{equation}
in units wherein $\hbar = m = c = 1$. Here $e$ is the charge of the vector meson,
$\vec{\Pi} = \left(\Pi_1, \Pi_2, \Pi_3\right)$ comes from the minimum coupling
substitution $ \vec{P} \to \vec{\Pi} = \vec{P} - e \vec{A}$, with the gauge symmetric
potential $A_1 = - \frac{1}{2} By$, $A_2 = \frac{1}{2} Bx$, $A_3 = 0$, and $\vec{B} = (0,
0, B)$, while
\begin{equation}
  \Sigma_3 = \left(\begin{array}{cc}
       S_3 & 0 \\
       0 & S_3
       \end{array}\right), \qquad
  S_3 = \left(\begin{array}{ccc}
       1 & 0 & 0 \\
       0 & 0 & 0 \\
       0 & 0 & -1
       \end{array}\right).
\end{equation}
\par
%
%
By using Johnson-Lippmann arguments, one can distinguish the so-called perpendicular
and parallel parts of Eq.~(\ref{eq:KG}) and, through the connection of the former
with a one-dimensional oscillator, obtain the squared relativistic energies as
\begin{equation}
  E^2 = 1 + 2 \omega \left[n + \case{1}{2}(\lambda + 1) - s \right],  \label{eq:Esquared}
\end{equation}
where $\omega \equiv eB$, $n$ is the Landau-level quantum number, and $s$ refers to the
eigenvalues of $S_3$. In the standard formulation of the Kemmer equation corresponding
to $\lambda=0$, some energy eigenvalues become complex for $\omega > 1$~\cite{vija}.
On the contrary, for $\lambda \ge 1$, all the energies remain real for any $\omega$. For
such $\lambda$ values, it can also be checked that the causality principle is
fulfilled~\cite{beckers95a}.\par
%
%
In the nonrelativistic limit, the perpendicular part of the Hamiltonian corresponding to
(\ref{eq:Esquared}) is given by
\begin{equation}
  H_{\rm NR}^{\bot} = \case{1}{2} \left(\Pi_1^2 + \Pi_2^2\right) + \case{1}{2}
  \omega (\I - 2S_3),
\end{equation}
where \I\ denotes the $3 \times 3$ unit matrix and the simple choice $\lambda = 1$ has
been made.\par
%
%
{}For $H_{\rm NR}^{\bot}$, one can construct two charge operators $Q_1$, $Q_2$, defined
by~\cite{beckers95a, beckers95b}
\begin{equation}
  Q_1 = {\cal A} \Pi_1 + {\cal B} \Pi_2, \qquad  Q_2 = - {\cal B} \Pi_1 + {\cal A} \Pi_2,
\end{equation}
where $\cal A$ and $\cal B$ are $3 \times 3$ odd matrices given by
\begin{equation}
  {\cal A} = \frac{1}{2\sqrt{2}} \left(\begin{array}{ccc}
      0 & 0 & 1+{\rm i} \\
      0 & 0 & -1+{\rm i} \\
      1-{\rm i} & -1-{\rm i} & 0
      \end{array}\right), \qquad
  {\cal B} = \frac{1}{2\sqrt{2}} \left(\begin{array}{ccc}
      0 & 0 & 1-{\rm i} \\
      0 & 0 & 1+{\rm i} \\
      1+{\rm i} & 1-{\rm i} & 0
      \end{array}\right).
\end{equation}
They satisfy the relations
\begin{eqnarray}
  Q_i^3 & = & Q_i H_{\rm NR}^{\bot}, \qquad \left[H_{\rm NR}^{\bot}, Q_i\right] = 0,
       \qquad i=1, 2, \label{eq:charges-1} \\
  Q_i^2 Q_j & = & Q_j Q_i^2 = - Q_i Q_j Q_i = Q_j H_{\rm NR}^{\bot}, \qquad i, j = 1, 2,
       \qquad i\ne j,  \label{eq:charges-2}
\end{eqnarray}
which differ from those characterizing either SSQM~\cite{witten81} or
PSSQM~\cite{rubakov, beckers90}. The charges $Q_1$, $Q_2$ correspond to the
superposition of usual bosons (associated with the even operators $\Pi_1$ and $\Pi_2$)
and {\em pseudofermions} (associated with the odd matrices $\cal A$ and $\cal B$).\par
%
%
In terms of linear combinations of the type
\begin{equation}
  Q = c (Q_1 - {\rm i}Q_2), \qquad Q^{\dagger} = c (Q_1 + {\rm i}Q_2), 
\end{equation}
where $c$ is some real constant (not to be confused with the velocity of light), the
algebra defined in (\ref{eq:charges-1}) and (\ref{eq:charges-2}) takes the form
\begin{equation}
  Q^2 = 0, \qquad \left[H_{\rm NR}^{\bot}, Q\right] = 0, \qquad Q Q^{\dagger} Q = 4c^2 Q
  H_{\rm NR}^{\bot},  \label{eq:charges-bis}
\end{equation}
together with the Hermitian conjugate relations. The first two equations in
(\ref{eq:charges-bis}) are the same as those occurring in SSQM~\cite{witten81},
whereas the third one is rather similar to the multilinear relation valid in PSSQM of order
two. Actually, for $c=1$ or $c=1/2$, it is compatible with the multilinear relation
appearing in Rubakov-Spiridonov-Khare~\cite{rubakov} or
Beckers-Debergh~\cite{beckers90} version of PSSQM, respectively.\par
%
%
With the choice $c = 1/2$, one can rewrite $Q$, $Q^{\dagger}$, and $H_{\rm NR}^{\bot}$
as
\begin{equation}
  Q = \sqrt{\omega}\, b a^{\dagger}, \qquad Q^{\dagger} = \sqrt{\omega}\, b^{\dagger} a,
  \qquad H_{\rm NR}^{\bot} = \omega \left[\case{1}{2} \left\{a, a^{\dagger}\right\} \I +
  \case{1}{2} \diag(-1, 1, 3)\right],  \label{eq:H_NR}
\end{equation}
where
\begin{equation}
  a = \frac{1}{\sqrt{2\omega}} (\Pi_1 + {\rm i} \Pi_2), \qquad a^{\dagger} = 
  \frac{1}{\sqrt{2\omega}} (\Pi_1 - {\rm i} \Pi_2)  \label{eq:boson}
\end{equation}
are bosonic creation and annihilation operators, while
\begin{eqnarray}
  b & = & \frac{1}{\sqrt{2}} ({\cal A} + {\rm i} {\cal B}) = \frac{1}{2}
        \left(\begin{array}{ccc}
        0 & 0 & 1 + {\rm i}\\
        0 & 0 & - 1 + {\rm i}\\
        0 & 0 & 0 
        \end{array}\right), \nonumber \\
  \bp & = & \frac{1}{\sqrt{2}} ({\cal A} - {\rm i} {\cal B}) = \frac{1}{2}
\left(\begin{array}{ccc}
        0 & 0 & 0\\
        0 & 0 & 0\\
        1 - {\rm i} & - 1 - {\rm i} & 0 
        \end{array}\right)  \label{eq:pseudofermion-bis}
\end{eqnarray}
are pseudofermionic ones, satisfying the relations
\begin{equation}
  b^2 = (\bp)^2 = 0, \qquad b \bp b = b, \qquad \bp b \bp = \bp. 
  \label{eq:pseudofermion}
\end{equation}
Such operators describe small violations of the Pauli principle.\par
%
%
The new symmetry described by Eq.~(\ref{eq:charges-bis}) is termed
pseudosupersymmetry, while the Hamiltonian in Eq.~(\ref{eq:H_NR}) is considered
as the pseudosupersymmetric oscillator~\cite{beckers95a, beckers95b}.\par
%
%
\section{Pseudosupersymmetric Quantum Mechanics in Terms of Two Superpotentials}

\setcounter{equation}{0}

As reviewed in Sec.~2, PsSSQM is characterized by a pseudosupersymmetric Hamiltonian
$\cal H$ and pseudosupercharge operators $Q$, $\Qp$ satisfying the relations
\begin{eqnarray}
  Q^2 & = & 0,  \label{eq:PsSSQM-1} \\[0pt]
  [{\cal H}, Q] & = & 0,  \label{eq:PsSSQM-2} \\
  Q \Qp Q & = & 4 c^2 Q {\cal H},  \label{eq:PsSSQM-3}
\end{eqnarray} 
and their Hermitian conjugates, where $c$ is some real constant~\cite{beckers95b}.\par
%
%
One may now look for a realization of the two pseudosupercharges $Q$, $\Qp$ as
\begin{eqnarray}
  Q & = & \frac{c}{\sqrt{2}} \left(\begin{array}{ccc}
        0 & 0 & (1-{\rm i}) [P + {\rm i} W_1(x)] \\
        0 & 0 & (1+{\rm i}) [P + {\rm i} W_2(x)] \\
        0 & 0 & 0
        \end{array}\right), \label{eq:Q-gen}\\
  \Qp & = & \frac{c}{\sqrt{2}} \left(\begin{array}{ccc}
        0 & 0 & 0 \\
        0 & 0 & 0 \\
        (1+{\rm i}) [P - {\rm i} W_1(x)] & (1-{\rm i}) [P - {\rm i} W_2(x)]  & 0
        \end{array}\right),  \label{eq:Qp-gen}
\end{eqnarray}
where $P$ is the momentum operator ($P = - {\rm i} d/dx$) and $W_1(x)$, $W_2(x)$
are two superpotentials. It is obvious that with this choice the
nilpotency~(\ref{eq:PsSSQM-1}) is verified.\par
%
%
Equation (\ref{eq:PsSSQM-2}) is fulfilled if the pseudosupersymmetric Hamiltonian $\cal
H$ is realized by a $3 \times 3$ Hermitian matrix of the form
\begin{equation}
  {\cal H} = \left(\begin{array}{ccc}
         H_1 & H_4 & 0 \\
         H_4^{\dagger} & H_2 & 0 \\
         0 & 0 & H_3
         \end{array}\right), \qquad H_i = H_i^{\dagger}, \quad i = 1, 2, 3,  \label{eq:H-gen}
\end{equation}
where $H_1$, $H_2$, $H_3$, and $H_4$ are constrained by the conditions
\begin{eqnarray}
  H_1 (P + {\rm i} W_1) + {\rm i} H_4 (P + {\rm i} W_2) & = & (P + {\rm i} W_1) H_3, 
        \label{eq:constraint-1}\\
  - {\rm i} H_4^{\dagger} (P + {\rm i} W_1) + H_2 (P + {\rm i} W_2) & = & (P + {\rm i}
        W_2) H_3.  \label{eq:constraint-2}
\end{eqnarray}
\par
%
%
{}Finally, Eq.~(\ref{eq:PsSSQM-3}) determines $H_3$ in terms of the two superpotentials
$W_1$, $W_2$ as 
\begin{eqnarray}
  H_3 & = & \case{1}{4} [(P - {\rm i} W_1) (P + {\rm i} W_1) + (P - {\rm i} W_2)
       (P + {\rm i} W_2)] \nonumber \\
  & = & \case{1}{2} \left[P^2 + \case{1}{2} \left(W_1^2 + W_2^2 + W'_1 + W'_2\right)
       \right].  \label{eq:H_3}
\end{eqnarray}
\par
%
%
By combining (\ref{eq:H_3}) with (\ref{eq:constraint-1}) and~(\ref{eq:constraint-2}),
the latter are transformed into
\begin{eqnarray}
  && (P + {\rm i} W_1) [(P - {\rm i} W_1)(P + {\rm i} W_1) + (P - {\rm i} W_2)(P +
       {\rm i} W_2)] \nonumber \\ 
  && \mbox{} = 4H_1 (P + {\rm i} W_1) + 4{\rm i} H_4 (P + {\rm i} W_2), 
       \label{eq:constraint-1bis}\\
  && (P + {\rm i} W_2) [(P - {\rm i} W_1)(P + {\rm i} W_1) + (P - {\rm i} W_2)(P +
       {\rm i} W_2)] \nonumber \\ 
  && \mbox{} = - 4{\rm i} H_4^{\dagger} (P + {\rm i} W_1) + 4H_2 (P + {\rm i} W_2).
       \label{eq:constraint-2bis}
\end{eqnarray}
The problem left amounts to solving this system of equations in order to express $H_1$,
$H_2$, $H_4$ in terms of $W_1$ and $W_2$. For such a purpose, we shall distinguish
between the two cases $W_1 = W_2$ and $W_1 \ne W_2$.\par
%
%
\subsection{The case of equal superpotentials}

Whenever the two superpotentials are equal,
\begin{equation}
  W_1(x) = W_2(x) = W(x),  \label{eq:W} 
\end{equation}
Eqs.~(\ref{eq:constraint-1bis}) and~(\ref{eq:constraint-2bis}) assume a very simple form
\begin{equation}
  (P + {\rm i} W)(P - {\rm i} W) = 2 (H_1 + {\rm i} H_4) = 2(H_2 - {\rm i}
  H_4^{\dagger}).  \label{eq:constraint-equal}
\end{equation}
Since the left-hand side of Eq.~(\ref{eq:constraint-equal}) is Hermitian, we immediately
get
\begin{eqnarray}
  H_4^{\dagger} & = & - H_4,  \label{eq:H_4-equal} \\
  H_1 & = & H_2 = \case{1}{2}(P^2 + W^2 - W') - {\rm i} H_4. 
\label{eq:H_1-equal}
\end{eqnarray}
\par
%
%
In the special case~(\ref{eq:W}), Eq.~(\ref{eq:H_3}) becomes
\begin{equation}
  H_3 = \case{1}{2}(P^2 + W^2 + W'),  \label{eq:H_3-equal}
\end{equation}
so that Eq.~(\ref{eq:H_1-equal}) may also be written as
\begin{equation}
  H_1 = H_2 = H_3 - {\rm i}(H_4 - {\rm i} W').  \label{eq:H_1-equalbis}
\end{equation}
\par
%
%
We conclude that in the case of equal superpotentials, the general solution of
Eqs.~(\ref{eq:constraint-1bis}), (\ref{eq:constraint-2bis}) is given by
(\ref{eq:H_4-equal}), (\ref{eq:H_1-equal}) (or~(\ref{eq:H_1-equalbis})),
and~(\ref{eq:H_3-equal}). Hence $H_4$ remains undetermined except for its
antihermitian character. Beckers and Debergh~\cite{beckers95b} restricted themlselves
to the choice $H_4 = {\rm i} W'$, in which case the three Hamiltonians $H_1$, $H_2$,
and $H_3$ become identical.\par
%
%
Such a restriction is however not needed. For an arbitrary solution (\ref{eq:H_4-equal}) --
(\ref{eq:H_3-equal}), we indeed note that $\cal H$, as given by~(\ref{eq:H-gen}), can be
diagonalized through a unitary transformation
\begin{equation}
  U_1 = \left(\begin{array}{ccc}
        \frac{1-{\rm i}}{2} & - \frac{1+{\rm i}}{2} & 0 \\
        0 & 0 &1 \\
        \frac{1}{\sqrt{2}} & \frac{{\rm i}}{\sqrt{2}} & 0
        \end{array}\right).
\end{equation}
The equivalent pseudosupersymmetric Hamiltonian and charges are given by
\begin{eqnarray}
  {\cal H}' & \equiv & U_1 {\cal H} U_1^{\dagger} = \case{1}{2}(P^2 + W^2) \I +
        \case{1}{2} W' \diag(-1, 1, -1) \nonumber \\
  && \mbox{} - 2{\rm i} H_4 \diag(0, 0, 1), \label{eq:equiv-H} \\
  Q' & \equiv & U_1 Q U_1^{\dagger} = - {\rm i}c \sqrt{2} \left(\begin{array}{ccc}
        0 & P + {\rm i}W & 0 \\
        0 & 0 & 0 \\
        0 & 0 & 0
        \end{array}\right),  \label{eq:equiv-Q} \\ 
  Q'^{\dagger} & \equiv & U_1 Q^{\dagger} U_1^{\dagger} = {\rm i}c \sqrt{2}
        \left(\begin{array}{ccc}
        0 & 0 & 0 \\
        P - {\rm i}W & 0 & 0 \\
        0 & 0 & 0
        \end{array}\right),  \label{eq:equiv-Q+}
\end{eqnarray}
where $\I$ again denotes the $3 \times 3$ unit matrix.\par
%
%
The Hamiltonian ${\cal H}'$ can be alternatively written as
\begin{equation}
  {\cal H}' = \left(\begin{array}{cc}
       H_{SS} &\begin{array}{c}
               0 \\
               0
               \end{array}\\
       \begin{array}{cc}
               0 & 0
       \end{array} & H_0
  \end{array}\right),  \label{eq:equiv-Hbis}
\end{equation}
where 
\begin{equation}
  H_{SS} = \left(\begin{array}{cc}
       \frac{1}{2}(P^2 + W^2 - W') & 0 \\
       0 & \frac{1}{2}(P^2 + W^2 + W') 
       \end{array}\right)
\end{equation}
is a standard supersymmetric Hamiltonian corresponding to the superpotential $W(x)$ and
\begin{equation}
  H_0 = H_1 - {\rm i}H_4 = H_3 - W' - 2{\rm i}H_4
\end{equation}
may be any Hermitian operator due to the arbitrariness of $H_4$.\par
%
%
{}From~(\ref{eq:equiv-Q}) and~(\ref{eq:equiv-Q+}), it is clear that the
pseudosupercharges $Q'$, $Q^{\prime\dagger}$ do not connect the eigenstates of
$H_0$ with those of $H_{SS}$. We also note that they may be related to the
orthosupercharges $Q_1$, $Q_1^{\dagger}$ of Khare {\em et al.}~\cite{khare} in
OSSQM of order two. We shall come back to the connections between PsSSQM and
OSSQM in Sec.~4.\par
%
%
In the special case $H_4 = {\rm i}W'$ considered by Beckers and
Debergh~\cite{beckers95b}, Eq.~(\ref{eq:equiv-H}) becomes
\begin{equation}
  {\cal H}' = \case{1}{2}(P^2 + W^2) \I + \case{1}{2} W' \diag(-1, 1, 3).
\end{equation}
For the oscillator-like superpotential $W(x) = x$ (in units wherein $\omega = 1$), one
then gets the pseudosupersymmetric oscillator Hamiltonian
\begin{equation}
  {\cal H}^{(1)\prime}_{\rm osc} = \case{1}{2} (P^2 + x^2) \I +
\case{1}{2} \diag(-1,
  1, 3),  \label{eq:H1-osc} 
\end{equation}
already encountered in Eq.~(\ref{eq:H_NR}) and whose spectrum contains the eigenvalues
$n$, $n+1$, and $n+2$ for $n=0$, 1, 2,~\ldots. The corresponding pseudosupercharges are 
\begin{equation}
  Q^{(1)\prime}_{\rm osc} = - {\rm i}c \sqrt{2} \left(\begin{array}{ccc}
        0 & P + {\rm i}x & 0 \\
        0 & 0 & 0 \\
        0 & 0 & 0
        \end{array}\right),  \qquad
  \left(Q^{(1)\prime}_{\rm osc}\right)^{\dagger} = {\rm i}c \sqrt{2}
        \left(\begin{array}{ccc}
        0 & 0 & 0 \\
        P - {\rm i}x & 0 & 0 \\
        0 & 0 & 0
        \end{array}\right).  \label{eq:Q1-osc} 
\end{equation}
The choice of~(\ref{eq:H1-osc}) in Ref.~\cite{beckers95b} to describe the
pseudosupersymmetric oscillator was dictated by the characteristics of the physical
problem at hand, namely that of relativistic vector mesons in a constant magnetic
field, as reviewed in Sec.~2.\par
%
%
If we disregard the application to such a problem, a natural choice for the
pseudosupersymmetric oscillator Hamiltonian and corresponding pseudosupercharges
would be
\begin{equation}
  {\cal H}^{(2)}_{\rm osc} = \case{1}{2} \{\ap, a\} \I + \case{1}{2} [\bp, b], \qquad
  Q^{(2)}_{\rm osc} = 2cb\ap, \qquad \left(Q^{(2)}_{\rm osc}\right)^{\dagger} = 2c
  \bp a,
\end{equation}
where $\ap = (x - {\rm i}P)/\sqrt{2}$, $a = (x + {\rm i}P)/\sqrt{2}$ are bosonic
creation and annihilation operators, $\bp$, $b$ are the pseudofermionic operators of
Eqs.~(\ref{eq:pseudofermion-bis}), (\ref{eq:pseudofermion}),  and both types of
operators are assumed to commute with one another. This choice corresponds to $W(x)
= x$, $H_4 = {\rm i}/4$ in Eqs. (\ref{eq:H_4-equal}) -- (\ref{eq:H_3-equal}) and leads to
the diagonalized pseudosupersymmetric Hamiltonian
\begin{equation}
  {\cal H}^{(2)\prime}_{\rm osc} = \case{1}{2}(P^2 + x^2) \I +
\case{1}{2} \diag(-1, 1,
  0),
\end{equation}
while the corresponding pseudosupercharges $Q^{(2)\prime}_{\rm osc}$,
$\left(Q^{(2)\prime}_{\rm osc}\right)^{\dagger}$ coincide with 
$Q^{(1)\prime}_{\rm osc}$, $\left(Q^{(1)\prime}_{\rm osc}\right)^{\dagger}$, given
in~(\ref{eq:Q1-osc}). The spectrum of ${\cal H}^{(2)\prime}_{\rm osc}$ contains the
eigenvalues $n$, $n+1$, and $n+\frac{1}{2}$ for $n=0$, 1, 2,~\ldots. Contrary to
what happens for ${\cal H}^{(1)\prime}_{\rm osc}$ whose levels starting from the
second excited state are threefold degenerate, those of ${\cal H}^{(2)\prime}_{\rm
osc}$ are either nondegenerate or twofold degenerate. This type of spectrum is
characteristic of the generic case for the Hamiltonian~(\ref{eq:equiv-H})
or~(\ref{eq:equiv-Hbis}) in view of the arbitrariness of $H_4$ or $H_0$.\par
%
%
Still another interesting special case corresponds to $H_4 = 0$ and
$W(x) = -x$, for which we get
\begin{eqnarray}
  {\cal H}^{(3)}_{\rm osc} & = & \case{1}{2}(P^2 + x^2) \I + \case{1}{2}
  \diag(1, 1, -1), \\
  {\cal H}^{(3)\prime}_{\rm osc}  & = & \case{1}{2}
  (P^2 + x^2) \I + \case{1}{2} \diag(1, -1, 1), \\
  Q^{(3)\prime}_{\rm osc} & = & - {\rm i}c \sqrt{2}
       \left(\begin{array}{ccc}
            0 & P - {\rm i}x & 0 \\
            0 & 0 & 0 \\
            0 & 0 & 0
       \end{array}\right),  \\
  \left(Q^{(3)\prime}_{\rm osc}\right)^{\dagger} & = & {\rm i}c
       \sqrt{2} \left(\begin{array}{ccc}
            0 & 0 & 0 \\
            P + {\rm i}x & 0 & 0 \\
            0 & 0 & 0
       \end{array}\right).  
\end{eqnarray}
The Hamiltonian ${\cal H}^{(3)\prime}_{\rm osc}$ coincides with that
of the Beckers-Debergh parasupersymmetric
oscillator~\cite{beckers90} and its spectrum contains the
eigenvalues $n+1$, $n$, and $n+1$ for
$n=0$, 1, 2,~\ldots. The corresponding charges are of course
different from the Beckers-Debergh parasupercharges.\par
%
%
\subsection{The case of unequal superpotentials}

Whenever the superpotentials are unequal, we are left with the
general equations (\ref{eq:constraint-1bis}),
(\ref{eq:constraint-2bis}). To solve them, let us write the first two
diagonal matrix elements of $\cal H$ as the sum of a kinetic and a
potential energy terms,
\begin{equation}
  H_i = \case{1}{2} P^2 + V_i(x), \qquad i=1, 2.  \label{eq:H_i}
\end{equation}
On inserting these expressions into Eqs.~(\ref{eq:constraint-1bis})
and~(\ref{eq:constraint-2bis}), writing the latter in normal order
form with $P$ on the right of $x$, and equating the coefficients of
equal powers of $P$, we obtain that $H_4$ does not depend on $P$
and that $V_1$, $V_2$, $H_4$ are related to $W_1$ and $W_2$
through the consistency conditions
\begin{eqnarray}
  V_1 + {\rm i}H_4 & = & \case{1}{4}(W_1^2 + W_2^2 - 3W'_1 + W'_2),
       \label{eq:consistency-1}\\
  V_1 W_1 + {\rm i}H_4 W_2 & = & \case{1}{4}(W_1^3 + W_1 W_2^2
       - W_1 W'_1 + W_1 W'_2 - 2W_2 W'_2 + W''_1 \nonumber \\
  && \mbox{} - W''_2), \\
  V_2 - {\rm i}H_4^{\dagger} & = & \case{1}{4}(W_1^2 + W_2^2 +
       W'_1 - 3W'_2), \\
  V_2 W_2 - {\rm i}H_4^{\dagger} W_1 & = & \case{1}{4}(W_1^2 W_2
       + W_2^3 - 2W_1 W'_1 + W_2 W'_1 - W_2 W'_2 - W''_1
       \nonumber \\
  && \mbox{} + W''_2).  \label{eq:consistency-4}
\end{eqnarray}
\par
%
%
Since the right-hand sides of these equations are real functions of
$x$, the same must be true for their left-hand sides. This implies
that $H_4$ must be an imaginary function of $x$, hence it satisfies
Eq.(\ref{eq:H_4-equal}) again. When taking the latter into account,
the set of Eqs.~(\ref{eq:consistency-1}) -- (\ref{eq:consistency-4})
becomes a nonhomogeneous system of four linear equations in three
unknowns $V_1$, $V_2$, and $H_4$. Such a system turns out to be
compatible and for $W_1 \ne W_2$ its solution is given by
\begin{eqnarray}
  V_1 & = & \frac{1}{4} \left[W_1^2 + W_2^2 - \frac{W_1-3W_2}
       {W_1-W_2} (W'_1 - W'_2) + \frac{W''_1-W''_2}{W_1-W_2}\right],
       \label{eq:V_1}\\
  V_2 & = & \frac{1}{4} \left[W_1^2 + W_2^2 + \frac{3W_1-W_2}
       {W_1-W_2} (W'_1 - W'_2) + \frac{W''_1-W''_2}{W_1-W_2}\right],
       \label{eq:V_2}\\
  H_4 & = & {\rm i} \frac{2W_1W'_1 - 2W_2W'_2 + W''_1 - W''_2}
       {4(W_1-W_2)}.  \label{eq:H_4-unequal}
\end{eqnarray}
\par
%
%
Before studying the general solution (\ref{eq:H_3}), (\ref{eq:H_i}),
(\ref{eq:V_1}) -- (\ref{eq:H_4-unequal}) in more detail, it is worth
considering the special case of equal and opposite superpotentials,
$W_1(x) = - W_2(x) = W(x)$~\cite{nicolas}. The solution then reduces
to
\begin{eqnarray}
  H_1 & = & \frac{1}{2} \left(P^2 + W^2 - 2W' + \frac{W''}{2W}\right), \\
  H_2  & = & \frac{1}{2} \left(P^2 + W^2 + 2W' +
       \frac{W''}{2W}\right), \\
  H_3 & = & \frac{1}{2} (P^2 + W^2), \quad H_4 = {\rm i}
       \frac{W''}{4W}. 
\end{eqnarray}
\par
%
%
The corresponding pseudosupersymmetric Hamiltonian $\cal H$
cannot be diagonalized through the unitary transformation $U_1$
anymore, but we note that for the oscillator-like superpotential
$W(x) = x$, $\cal H$ is diagonal so that we still get another
pseudosupersymmetric oscillator Hamiltonian
\begin{equation}
  {\cal H}^{(4)}_{\rm osc} = \case{1}{2}(P^2 + x^2) \I + \diag(-1, 1, 0).
  \label{eq:H4}
\end{equation}
By a permutation of rows and columns corresponding to the unitary
matrix
\begin{equation}
  U_2 = \left(\begin{array}{ccc}
       1 & 0 & 0 \\
       0 & 0 & 1 \\
       0 & 1 & 0
       \end{array}\right),
\end{equation}
the latter may be transformed into the Rubakov-Spiridonov-Khare
parasupersymmetric oscillator Hamiltonian~\cite{rubakov},
\begin{equation}
  {\cal H}^{(4)\prime}_{\rm osc} \equiv U_2 {\cal H}^{(4)}_{\rm osc}
  U_2^{\dagger} = \case{1}{2}(P^2 + x^2) \I + \diag(-1, 0, 1),  
  \label{eq:H4bis}
\end{equation}
whose spectrum contains the eigenvalues $n-\frac{1}{2}$,
$n+\frac{1}{2}$, and $n+\frac{3}{2}$ for $n=0$, 1, 2,~\ldots.\par
%
%
The same transformation leads to the pseudosupercharge
\begin{equation}
  Q^{(4)\prime}_{\rm osc} \equiv U_2 Q^{(4)}_{\rm osc} U_2^{\dagger}
  = \frac{c}{\sqrt{2}} \left(\begin{array}{ccc}
       0 & (1-{\rm i})(P + {\rm i}x) & 0 \\
       0 & 0 & 0 \\
       0 & (1+{\rm i})(P - {\rm i}x) & 0
  \end{array}\right)  \label{eq:Q4}
\end{equation}
and its Hermitian conjugate, which of course differ from
Rubakov-Spiridonov-Khare parasupercharges. Contrary to the
pseudosupercharges (\ref{eq:equiv-Q}), (\ref{eq:equiv-Q+}), those
given in~(\ref{eq:Q4}) and its Hermitian conjugate connect the
eigenstates of all the components of the pseudosupersymmetric
Hamiltonian. Hence the model described by (\ref{eq:H4}) or (\ref{eq:H4bis}) cannot be
split into two isospectral models, one scalar and one $2 \times 2$, as it was the
case for that corresponding to (\ref{eq:H1-osc}).\par
%
%
Going back now to the general solution for $W_1 \ne W_2$, we may
ask which choice of superpotentials makes $\cal H$ diagonal. From
Eq.~(\ref{eq:H_4-unequal}), we immediately see that the condition $H_4
= 0$ leads to the differential equation
\begin{equation}
  W''_1 + 2W_1 W'_1 = W''_2 + 2W_2 W'_2,
\end{equation}
which is equivalent to
\begin{equation}
  W'_1 + W_1^2 = W'_2 + W_2^2 + C,  \label{eq:cond-W}
\end{equation}
where $C$ is some real integration constant. By replacing $W_1$ and
$W_2$ by the linear combinations $W_{\pm} = W_1 \pm W_2$,
Eq.~(\ref{eq:cond-W}) is transformed into
\begin{equation}
  W'_- + W_+ W_- = C,
\end{equation}
which can be easily solved to yield $W_-$ in terms of $W_+$. The
results for $W_1$ and $W_2$ read
\begin{eqnarray}
  W_1(x) & = & \frac{1}{2} \left(W_+(x) + \frac{C \int^x \exp\left(
       \int^t W_+(u) du\right) dt + D}{\exp\left(\int^x W_+(t) dt\right)}
       \right), \label{eq:W_1-diag} \\
  W_2(x) & = & \frac{1}{2} \left(W_+(x) - \frac{C \int^x \exp\left(
       \int^t W_+(u) du\right) dt + D}{\exp\left(\int^x W_+(t) dt\right)}
       \right),  \label{eq:W_2-diag}
\end{eqnarray}
where $D$ is another real integration constant. We conclude that for
any real function $W_+(x)$ and any real constants $C$, $D$, the
choice (\ref{eq:W_1-diag}), (\ref{eq:W_2-diag}) for $W_1$ and
$W_2$ ensures that $\cal H$ is diagonal.\par
%
%
The simplest choice for $W_+$ is $W_+ = 0$. We then obtain from
(\ref{eq:W_1-diag}) and (\ref{eq:W_2-diag}) that $W_1(x) = - W_2(x)
= \frac{1}{2}(Cx+D)$, which for $C=2$ and $D=0$ leads to the
pseudosupersymmetric Hamiltonian ${\cal H}^{(4)}_{\rm osc}$ given
in Eq.~(\ref{eq:H4}).\par
%
%
\section{Connection with Orthosupersymmetric Quantum Mechanics}

\setcounter{equation}{0}

In Sec.~3, we have already established some connections between
PsSSQM and PSSQM for oscillator-like superpotentials. In the
present section, we turn ourselves to OSSQM and study its
relationship with PsSSQM.\par
%
%
To start with, we note that pseudofermion operators can be
constructed in terms of orthofermion creation and annihilation
operators of order two $\cp_{\alpha}$, $c_{\alpha}$, $\alpha=1$, 2,
defined by the relations
\begin{equation}
  c_{\alpha} c_{\beta} = 0, \qquad c_{\alpha} \cp_{\beta} +
  \delta_{\alpha,\beta} \sum_{\gamma=1}^2 \cp_{\gamma}
  c_{\gamma} = \delta_{\alpha,\beta}, \qquad \alpha, \beta = 1, 2, 
  \label{eq:orthofermion}
\end{equation}
and their Hermitian conjugates~\cite{mishra}. It is indeed clear that
the linear combinations with complex coefficients
\begin{equation}
  \bt = \xi \cp_1 + \eta \cp_2, \qquad \btp = \xi^* c_1 + \eta^* c_2,
  \qquad |\xi|^2 + |\eta|^2 = 1,  \label{eq:bt}
\end{equation}
satisfy Eq.~(\ref{eq:pseudofermion}).\par
%
%
In the standard three-dimensional matrix representation of the
orthofermion algebra, wherein
\begin{equation}
  c_1 = \left(\begin{array}{ccc}
       0 & 1 & 0 \\
       0 & 0 & 0 \\
       0 & 0 & 0
       \end{array}\right), \quad 
  c_2 = \left(\begin{array}{ccc}
       0 & 0 & 1 \\
       0 & 0 & 0 \\
       0 & 0 & 0
       \end{array}\right), \quad 
  \cp_1 = \left(\begin{array}{ccc}
       0 & 0 & 0 \\
       1 & 0 & 0 \\
       0 & 0 & 0
       \end{array}\right), \quad 
  \cp_2 = \left(\begin{array}{ccc}
       0 & 0 & 0 \\
       0 & 0 & 0 \\
       1 & 0 & 0
       \end{array}\right), 
\end{equation}
$\bt$ and $\btp$ are represented by
\begin{equation}
  \bt = \left(\begin{array}{ccc}
       0 & 0 & 0 \\
       \xi & 0 & 0 \\
       \eta & 0 & 0
       \end{array}\right), \qquad
  \btp = \left(\begin{array}{ccc}
       0 & \xi^* & \eta^* \\
       0 & 0 & 0 \\
       0 & 0 & 0
       \end{array}\right).
\end{equation}
Such matrices are unitarily equivalent to the standard matrix
realization (\ref{eq:pseudofermion-bis}) of the pseudofermionic
operators $b$, $\bp$ since the matrices
\begin{equation}
  b \equiv U_3 \bt U_3^{\dagger} = \left(\begin{array}{ccc}
       0 & 0 & \xi \\
       0 & 0 & \eta \\
       0 & 0 & 0
       \end{array}\right), \qquad
  \bp \equiv U_3 \btp U_3^{\dagger} = \left(\begin{array}{ccc}
       0 & 0 & 0 \\
       0 & 0 & 0 \\
       \xi^* & \eta^* & 0
       \end{array}\right), 
\end{equation}
where
\begin{equation}
  U_3 = \left(\begin{array}{ccc}
       0 & 1 & 0 \\
       0 & 0 & 1 \\
       1 & 0 & 0
       \end{array}\right),  \label{eq:U_3}
\end{equation}
coincide with (\ref{eq:pseudofermion-bis}) provided we make the
choice
\begin{equation}
  \xi = \case{1}{2}(1 + {\rm i}), \qquad \eta = \case{1}{2}(- 1 + {\rm
  i})  \label{eq:xi-eta}
\end{equation}
in (\ref{eq:bt}).\par
%
%
Let us now consider an orthosupersymmetric Hamiltonian ${\cal
H}^K$ of order two, satisfying the relations
\begin{eqnarray}
  && Q^K_{\alpha} Q^K_{\beta} = 0, \label{eq:OSSQM-1}\\
  && [{\cal H}^K, Q^K_{\alpha}] = 0, \\
  && Q^K_{\alpha} \left(Q^K_{\beta}\right)^{\dagger} + \delta_{\alpha,\beta}
       \sum_{\gamma=1}^2 \left(Q^K_{\gamma}\right)^{\dagger} Q^K_{\gamma} =
       2 \delta_{\alpha,\beta} {\cal H}^K,  \label{eq:OSSQM-3}  
\end{eqnarray}
where $Q^K_{\alpha}$, $\left(Q^K_{\alpha}\right)^{\dagger}$, $\alpha=1$, 2, are the
orthosupercharge operators~\cite{khare}. The above relationship between orthofermions
of order two and pseudofermions suggests the construction of the operators 
\begin{equation}
  \Qt = \zeta \left(Q^K_1\right)^{\dagger} + \rho \left(Q^K_2\right)^{\dagger},
  \quad \Qtp = \zeta^* Q^K_1 + \rho^* Q^K_2, \quad \Ht = {\cal H}^K, \quad
  |\zeta|^2 + |\rho|^2 = 2c^2,  \label{eq:Qt}  
\end{equation}
which can be checked to satisfy the defining relations (\ref{eq:PsSSQM-1}) --
(\ref{eq:PsSSQM-3}) of PsSSQM. We conclude that any order-two orthosupersymmetric
quantum mechanical system has a pseudosupersymmetry generated by $\Qt$ and
$\Qtp$, given in~(\ref{eq:Qt}).\par
%
%
Both properties (\ref{eq:bt}) and (\ref{eq:Qt}) can actually be extended to order-$p$
orthofermionic operators and order-$p$ orthosupersymmetric quantum mechanical
systems, respectively. The corresponding equations read
\begin{equation}
  \bt = \sum_{\alpha=1}^p \xi_{\alpha} \cp_{\alpha}, \qquad \btp =
  \sum_{\alpha=1}^p \xi^*_{\alpha} c_{\alpha}, \qquad \sum_{\alpha=1}^p
  |\xi_{\alpha}|^2 = 1, 
\end{equation}
and
\begin{equation}
  \Qt = \sum_{\alpha=1}^p \zeta_{\alpha} \left(Q^K_{\alpha}\right)^{\dagger}, \qquad
  \Qtp = \sum_{\alpha=1}^p \zeta^*_{\alpha} Q^K_{\alpha}, \qquad \Ht = {\cal H}^K,
  \qquad \sum_{\alpha=1}^p |\zeta_{\alpha}|^2 = 2c^2, 
\end{equation}
where $c_{\alpha}$, $\cp_{\alpha}$ and $Q^K_{\alpha}$,
$\left(Q^K_{\alpha}\right)^{\dagger}$, ${\cal H}^K$ satisfy equations similar to
(\ref{eq:orthofermion}) and (\ref{eq:OSSQM-1}) -- (\ref{eq:OSSQM-3}) with all indices
$\alpha$, $\beta$, $\gamma$ running from 1 to $p$.\par
%
%
To study in more detail the connections between order-two OSSQM and PsSSQM, let us
consider the OSSQM realization of Khare {\em et al.} in terms of two superpotentials
$W^K_1(x)$, $W^K_2(x)$~\cite{khare} and compare it with the corresponding
realization of PsSSQM, given in Sec.~3. On using the former, we get the following
realization for the operators of Eq.~(\ref{eq:Qt}),
\begin{equation}
  \Qt = \left(\begin{array}{ccc}
       0 & 0 & 0 \\
       \zeta (P + {\rm i}W^K_1) & 0 & 0 \\
       \rho (P + {\rm i}W^K_2) & 0 & 0
       \end{array}\right), \qquad
  \Qtp = \left(\begin{array}{ccc}
       0 & \zeta^* (P - {\rm i}W^K_1) & \rho^* (P - {\rm i}W^K_2) \\
       0 & 0 & 0 \\
       0 & 0 & 0
       \end{array}\right), 
\end{equation}
\begin{equation}
  \Ht = \left(\begin{array}{ccc}
       h_1 & 0 & 0 \\
       0 & h_2 & 0 \\
       0 & 0 & h_3
       \end{array}\right),
\end{equation}
where
\begin{eqnarray}
  h_1 & = & \case{1}{2} \left[p^2 + (W^K_1)^2 + (W^K_1)'\right], \label{eq:h_1} \\
  h_2 & = & \case{1}{2} \left[p^2 + (W^K_1)^2 - (W^K_1)'\right], \label{eq:h_2}\\
  h_3 & = & \case{1}{2} \left[p^2 + (W^K_2)^2 - (W^K_2)'\right],  \label{eq:h_3} 
\end{eqnarray}
and $W^K_1$, $W^K_2$ are constrained by the relation
\begin{equation}
  (W^K_1)^2 + (W^K_1)' = (W^K_2)^2 + (W^K_2)'.  \label{eq:OSSQM-constraint}
\end{equation}
Applying now the unitary transformation (\ref{eq:U_3}), we obtain the matrices
\begin{eqnarray}
  Q & \equiv & U_3 \Qt U_3^{\dagger} = \left(\begin{array}{ccc}
       0 & 0 & \zeta (P + {\rm i}W^K_1) \\
       0 & 0 & \rho (P + {\rm i}W^K_2) \\
       0 & 0 & 0
       \end{array}\right), \label{eq:Q-OSSQM}\\
  \Qp & \equiv & U_3 \Qtp U_3^{\dagger} = \left(\begin{array}{ccc}
       0 & 0 & 0 \\
       0 & 0 & 0 \\
       \zeta^* (P - {\rm i}W^K_1) & \rho^* (P - {\rm i}W^K_2) & 0
       \end{array}\right), \label{eq:Qp-OSSQM}\\
  {\cal H} & \equiv & U_3 \Ht U_3^{\dagger} = \left(\begin{array}{ccc}
       h_2 & 0 & 0 \\
       0 & h_3 & 0 \\
       0 & 0 & h_1
       \end{array}\right),  \label{eq:H-OSSQM} 
\end{eqnarray}
which have to be compared with the standard realization of PsSSQM in terms of two
superpotentials $W_1$, $W_2$, given in (\ref{eq:Q-gen}) -- (\ref{eq:H-gen}).\par
%
%
More precisely, we would like to determine under which conditions a PsSSQM system
described by Eqs.~(\ref{eq:Q-gen}) -- (\ref{eq:H-gen}) may be characterized by
Eqs.~(\ref{eq:Q-OSSQM}) -- (\ref{eq:H-OSSQM}) constructed from a matrix realization of
OSSQM. Comparison between Eqs.~(\ref{eq:Q-gen}), (\ref{eq:Qp-gen}) and
(\ref{eq:Q-OSSQM}), (\ref{eq:Qp-OSSQM}) directly leads to the identifications
\begin{equation}
  \zeta = - {\rm i}c\sqrt{2} \xi = \frac{c(1-{\rm i})}{\sqrt{2}}, \qquad \rho = - {\rm
  i}c\sqrt{2} \eta = \frac{c(1+{\rm i})}{\sqrt{2}}, 
\end{equation}
and
\begin{equation}
  W^K_1(x) = W_1(x), \qquad W^K_2(x) = W_2(x),  \label{eq:O-Ps-1}
\end{equation}
where we have taken Eq.~(\ref{eq:xi-eta}) into account. Furthermore,
Eqs.~(\ref{eq:H-gen}) and (\ref{eq:H-OSSQM}) give rise to the constraints
\begin{equation}
  h_1 = H_3, \qquad h_2 = H_1, \qquad h_3 = H_2, \qquad H_4 = 0,
  \label{eq:O-Ps-2}
\end{equation}
which have to be compatible with the expressions of $h_1$, $h_2$, $h_3$ and $H_1$,
$H_2$, $H_3$, $H_4$ in terms of $W^K_1$, $W^K_2$ and $W_1$ ,$W_2$,
respectively.\par
%
%
In the case of equal superpotentials $W_1 = W_2 = W$ considered in Subsec.~3.1, we
note that Eq.~(\ref{eq:O-Ps-1}) implies that Eq.~(\ref{eq:OSSQM-constraint}) is
automatically satisfied. It then remains to impose the sole condition $H_4=0$, because if
the latter is satisfied, it results from Eqs.~(\ref{eq:H_1-equal}), (\ref{eq:H_3-equal}),
and (\ref{eq:h_1}) -- (\ref{eq:h_3}) that the same is true for the
remaining constraints in Eq.~(\ref{eq:O-Ps-2}).\par
%
%
In the case of unequal superpotentials $W_1 \ne W_2$ discussed in Subsec.~3.2, we
know that the condition $H_4=0$ leads to the constraint (\ref{eq:cond-W}), where $C$
may be any real constant. On taking Eq.~(\ref{eq:O-Ps-1}) into account, it is clear that
the two constraints (\ref{eq:cond-W}) and (\ref{eq:OSSQM-constraint}) are compatible
only if $C=0$. It is then straightforward to check that if we impose such a restriction, the
remaining conditions in Eq.~(\ref{eq:O-Ps-2}) are fulfilled by the operators
(\ref{eq:H_3}), (\ref{eq:H_i}), (\ref{eq:V_1}), (\ref{eq:V_2}) and (\ref{eq:h_1}) --
(\ref{eq:h_3}).\par
%
%
We conclude that any PsSSQM system for which $H_4=0$ may be considered as an
order-two OSSQM one with the same superpotentials provided either $W_1 = W_2$ or
$W_1 \ne W_2$ and the integration constant $C$ in Eq.~(\ref{eq:cond-W}) vanishes.
Order-two OSSQM systems are therefore a subclass of PsSSQM ones. Examples of PsSSQM
systems that cannot be considered as OSSQM ones are provided by the
pseudosupersymmetric oscillator Hamiltonians ${\cal H}^{(1)}_{\rm osc}$, ${\cal
H}^{(2)}_{\rm osc}$, and ${\cal H}^{(4)}_{\rm osc}$, defined in Sec.~3. The first two
correspond to $W_1 = W_2$ and $H_4 \ne 0$, while the third one is obtained for
$W_1 \ne W_2$ and $C=2$. Such a Hamiltonian being unitarily equivalent to the
Rubakov-Spiridonov-Khare parasupersymmetric oscillator has a negative-energy ground
state, whereas such a phenomenon cannot occur in OSSQM.\par
%
%
\section{Pseudosupersymmetric Quantum Mechanics in Terms of Generalized Deformed
Oscillator Algebra Generators}

\setcounter{equation}{0}

The purpose of the present section is to propose two new realizations of PsSSQM in terms
of the generators of a GDOA. The latter may be defined as a nonlinear associative algebra 
\AG\ generated by the operators $N = N^{\dagger}$, $\ap$, and $a = (\ap)^{\dagger}$,
satisfying the commutation relations
\begin{equation}
  [N, \ap] = \ap, \qquad [N, a] = - a, \qquad [a, \ap] = G(N),  \label{eq:GDOA}
\end{equation}
where $G(N) = [G(N)]^{\dagger}$ is some Hermitian function of $N$~\cite{katriel}.\par
%
%
We restrict ourselves here to GDOAs possessing a bosonic Fock space representation. In
the latter, we may write
\begin{equation}
  \ap a = F(N), \qquad a \ap = F(N+1),
\end{equation}
where the structure function $F(N) = [F(N)]^{\dagger}$ is such that
\begin{equation}
  G(N) = F(N+1) - F(N)  \label{eq:G-F}
\end{equation}
and is assumed to satisfy the conditions $F(0) = 0$ and $F(n) > 0$ if $n=1$, 2,
3,~\ldots. The carrier space $\cal F$ of such a representation can be constructed from a
vacuum state $|0\rangle$ (such that $a |0\rangle = N |0\rangle = 0$) by successive
applications of the creation operator $\ap$. Its basis states
\begin{equation}
  |n\rangle = \left(\prod_{i=1}^n F(i)\right)^{-1/2} (\ap)^n |0\rangle, \qquad n=0, 1, 2,
  \ldots,  \label{eq:basis}
\end{equation}
satisfy the relations
\begin{equation}
  N |n\rangle = n |n\rangle, \qquad \ap |n\rangle = \sqrt{F(n+1)} |n+1\rangle, \qquad
  a |n\rangle = \sqrt{F(n)} |n-1\rangle.
\end{equation}
\par
%
%
Note that for $G(N) = I$, $F(N) = N$, the algebra \AG\ reduces to the standard (bosonic)
oscillator algebra ${\cal A}(I)$, for which the creation and annihilation operators may be
written as $\ap = (x - {\rm i}P)/\sqrt{2}$, $a = (x + {\rm i}P)/\sqrt{2}$ or,
alternatively, as in Eq.~(\ref{eq:boson}).\par
%
%
Let us first consider the pseudosupercharges (\ref{eq:equiv-Q}), (\ref{eq:equiv-Q+}), and
the pseudosupersymmetric Hamiltonian (\ref{eq:equiv-Hbis}), obtained in the case of
equal superpotentials. For $W(x) = x$, they become
\begin{equation}
  Q' = 2c \left(\begin{array}{ccc}
       0 & \ap & 0 \\
       0 & 0 & 0 \\
       0 & 0 & 0
       \end{array}\right), \quad
  Q^{\prime\dagger} = 2c \left(\begin{array}{ccc}
       0 & 0 & 0 \\
       a & 0 & 0 \\
       0 & 0 & 0
       \end{array}\right), \quad
  {\cal H}' = \left(\begin{array}{ccc}
       \ap a & 0 & 0 \\
       0 & a \ap & 0 \\
       0 & 0 & H_0
       \end{array}\right),  \label{eq:Q'} 
\end{equation}
where $\ap$, $a$ are standard bosonic operators belonging to ${\cal A}(I)$. Inspired by
this remark and some results for SSQM~\cite{bonatsos}, let us introduce the matrices
\begin{eqnarray}
  \Qb & = & 2c \left(\begin{array}{ccc}
       0 & f(N) \ap & 0 \\
       0 & 0 & 0 \\
       0 & 0 & 0
       \end{array}\right), \qquad
  \Qbp = 2c \left(\begin{array}{ccc}
       0 & 0 & 0 \\
       f(N+1) a & 0 & 0 \\
       0 & 0 & 0
       \end{array}\right),  \label{eq:Qb-1} \\
  \bar{\cal H} & = & \left(\begin{array}{ccc}
       \Hb_1 & 0 & 0 \\
       0 & \Hb_2 & 0 \\
       0 & 0 & \Hb_3       
       \end{array}\right),   
\end{eqnarray}
where $N$, $\ap$, $a$ are the generators of some GDOA \AG, $f(N)$ is some real
function of $N$, and $\Hb_i$, $i=1$, 2, 3, are some $N$-dependent  Hermitian
operators. It is straightforward to show that such operator-valued matrices satisfy the
defining relations (\ref{eq:PsSSQM-1}) -- (\ref{eq:PsSSQM-3}) of PsSSQM provided we
choose
\begin{equation}
  \Hb_1 = f^2(N) F(N), \qquad \Hb_2 = f^2(N+1) F(N+1),  \label{eq:Hb-12}
\end{equation}
while $\Hb_3$ remains arbitrary. For $f(N) = 1$ and $F(N) = N$, we get back ${\cal
H}'$ in Eq.~(\ref{eq:Q'}) with $\Hb_3 = H_0$. For arbitrary $f(N)$ and $F(N)$, the
spectrum of $\bar{\cal H}$ contains the eigenvalues $f^2(n) F(n)$ and $f^2(n+1)
F(n+1)$, $n=0$, 1, 2, \ldots, as well as those of $\Hb_3$, and is clearly nonlinear. For
generic $\Hb_3$, the levels are either nondegenerate or twofold degenerate.\par
%
%
Let us next consider the pseudosupersymmetric Hamiltonian ${\cal H}^{(4)\prime}_{\rm
osc}$ and the pseudosupercharges $Q^{(4)\prime}_{\rm osc}$,
$\left(Q^{(4)\prime}_{\rm osc}\right)^{\dagger}$, corresponding to the choice $W_1(x)
= - W_2(x) = x$ and given in (\ref{eq:H4bis}) and (\ref{eq:Q4}), respectively. By a
procedure similar to that used above, we are led to propose another realization of PsSSQM,
\begin{eqnarray}
  \Qb & = & c\sqrt{2} \left(\begin{array}{ccc}
        0 & f_1(N) \ap & 0 \\
        0 & 0 & 0 \\
        0 & {\rm i} f_2(N+1) a & 0
        \end{array}\right), \label{eq:Qb-2}\\
  \Qbp & = & c\sqrt{2} \left(\begin{array}{ccc}
        0 & 0 & 0 \\
        f_1(N+1) a & 0 & -{\rm i} f_2(N) \ap \\
        0 & 0 & 0
        \end{array}\right),  \label{eq:Qbp-2} \\
  \bar{{\cal H}} & = & \left(\begin{array}{ccc}
       \Hb_1 & 0 & 0 \\
       0 & \Hb_2 & 0 \\
       0 & 0 & \Hb_3       
       \end{array}\right),  \label{eq:Hb-2} 
\end{eqnarray} 
where $N$, $\ap$, $a$ are the generators of \AG\ again, $f_1(N)$, $f_2(N)$ some real
functions of $N$, and $\Hb_i$, $i=1$, 2, 3, are some $N$-dependent  Hermitian
operators. This time we find that $\Hb_1$, $\Hb_2$, $\Hb_3$ are constrained by the
relations
\begin{eqnarray}
  \Hb_1 & = & \case{1}{2} [f_1^2(N) F(N) + f_2^2(N-1) F(N-1)], \\
  \Hb_2 & = & \case{1}{2} [f_1^2(N+1) F(N+1) + f_2^2(N) F(N)], \\
  \Hb_3 & = & \case{1}{2} [f_1^2(N+2) F(N+2) + f_2^2(N+1) F(N+1)].  \label{eq:Hb_3}
\end{eqnarray} 
For $f_1(N) = f_2(N) = 1$ and $F(N) = N$, $\bar{\cal H}$ reduces to ${\cal
H}^{(4)\prime}_{\rm osc} = \diag\left(N-\frac{1}{2}, N+\frac{1}{2},
N+\frac{3}{2}\right)$, while $\Qb$ and $\Qbp$ only differ from
$Q^{(4)\prime}_{\rm osc}$ and $\left(Q^{(4)\prime}_{\rm osc}\right)^{\dagger}$ by
an irrelevant phase factor. For arbitrary $f_1(N)$, $f_2(N)$, and $F(N)$, the spectrum
of $\bar{\cal H}$ is nonlinear and contains the eigenvalues $\frac{1}{2} [f_1^2(n) F(n)
+ f_2^2(n-1) F(n-1)]$, $\frac{1}{2} [f_1^2(n+1) F(n+1) + f_2^2(n) F(n)]$, and
$\frac{1}{2} [f_1^2(n+2) F(n+2) + f_2^2(n+1) F(n+1)]$ for $n=0$, 1, 2,~\ldots.
Except for a single nondegenerate level and a single twofold-degenerate one, all the
levels are threefold degenerate.\par
%
%
It is worth mentioning that the Hamiltonian $\bar{\cal H}$ of Eqs.~(\ref{eq:Hb-2}) --
(\ref{eq:Hb_3}) is also obtained with a different set of charges in a realization of
Rubakov-Spiridonov-Khare PSSQM in terms of \AG\ generators~\cite{cq02}.\par
%
%
\section{Reducibility and Bosonization of Pseudosupersymmetric Quantum Mechanics}

\setcounter{equation}{0}

We now plan to show that the two matrix realizations of PsSSQM in terms of GDOA
generators, defined in Eqs.~(\ref{eq:Qb-1}) -- (\ref{eq:Hb-12}) and (\ref{eq:Qb-2}) --
(\ref{eq:Hb_3}), respectively, are fully reducible.\par
%
%
{}For such a purpose, let us introduce the operators
\begin{equation}
  P_{\mu} = \frac{1}{3} \sum_{\nu=0}^2 e^{-2\pi{\rm i}\mu\nu/3} T^{\nu}, \qquad
  \mu=0, 1, 2,  \label{eq:P}
\end{equation}
where
\begin{equation}
  T = e^{2\pi{\rm i}N/3}.  \label{eq:T} 
\end{equation}
It is straightforward to show that in the Fock representation $T^3 = I$ and the
operators $P_{\mu}$ project on the subspaces ${\cal F}_{\mu} \equiv \{ |k\lambda
+ \mu \rangle \mid k=0, 1, 2, \ldots\}$ of the Fock space $\cal F$. Here $|n\rangle =
|k\lambda + \mu\rangle$ are the basis states~(\ref{eq:basis}). We actually obtain a
decomposition of $\cal F$ into three mutually orthogonal subspaces, ${\cal F} =
\sum_{\mu=0}^2 \oplus {\cal F}_{\mu}$. In other words, the operators $P_{\mu}$
satisfy the relations
\begin{equation}
  P_{\mu}^{\dagger} = P_{\mu}, \qquad P_{\mu} P_{\nu} = \delta_{\mu,\nu} P_{\mu},
  \qquad \sum_{\mu=0}^2 P_{\mu} = I  \label{eq:P-prop1}
\end{equation}
in $\cal F$. From Eqs.~(\ref{eq:GDOA}), (\ref{eq:P}), and (\ref{eq:T}), we also derive
the additional relations
\begin{equation}
  [N, T] = 0, \qquad \ap T = e^{-2\pi{\rm i}/3} T \ap,  \label{eq:T-prop} 
\end{equation}
and
\begin{equation}
  [N, P_{\mu}] = 0, \qquad \ap P_{\mu} = P_{\mu+1} \ap,  \label{eq:P-prop2}
\end{equation}
where we use the convention $P_{\mu'} = P_{\mu}$ if $\mu' - \mu = 0 \mod 3$.\par
%
%
Let us next consider the $3 \times 3$ matrix
\begin{equation}
  U_4 = \left(\begin{array}{ccc}
         P_0 & P_2 & P_1 \\
         P_1 & P_0 & P_2 \\
         P_2 & P_1 & P_0
         \end{array}\right),
\end{equation}
whose elements are $P_{\mu}$ operators. From (\ref{eq:P-prop1}), it results that 
\begin{equation}
  U_4^{\dagger} = \left(\begin{array}{ccc}
         P_0 & P_1 & P_2 \\
         P_2 & P_0 & P_1 \\
         P_1 & P_2 & P_0
         \end{array}\right) 
\end{equation}
and
\begin{equation}
  U_4 U_4^{\dagger} = U_4^{\dagger} U_4 = \I,
\end{equation}
showing that $U_4$ is a unitary matrix.\par
%
%
Through the unitary transformation represented by $U_4$, we then easily get diagonal
realizations of PsSSQM equivalent to (\ref{eq:Qb-1}) -- (\ref{eq:Hb-12}) and 
(\ref{eq:Qb-2}) -- (\ref{eq:Hb_3}),
\begin{eqnarray}
  \Qb' \equiv U_4 \Qb U_4^{\dagger} & = & \left(\begin{array}{ccc}
        \Qb_0 & 0 & 0 \\
        0 & \Qb_1 & 0 \\
        0 & 0 & \Qb_2
        \end{array}\right), \\
  \Qb^{\prime\dagger} \equiv U_4 \Qbp U_4^{\dagger} & = &
        \left(\begin{array}{ccc}
        \Qbp_0 & 0 & 0 \\
        0 & \Qbp_1 & 0 \\
        0 & 0 & \Qbp_2
        \end{array}\right),  \\
  \bar{\cal H}' \equiv U_4 \bar{\cal H} U_4^{\dagger} & = & \left(\begin{array}{ccc}
        \bar{\cal H}_0 & 0 & 0 \\
        0 & \bar{\cal H}_1 & 0 \\
        0 & 0 & \bar{\cal H}_2
        \end{array}\right), 
\end{eqnarray}
where $\Qb_{\mu}$, $\Qbp_{\mu}$, $\bar{\cal H}_{\mu}$, $\mu=0$, 1, 2, are given by
\begin{eqnarray}
  \Qb_{\mu} & = & 2c f(N) \ap P_{\mu+2}, \qquad \Qbp_{\mu} = 2c f(N+1) a P_{\mu},
        \label{eq:boson1-1}\\
  \bar{\cal H}_{\mu} & = & \sum_{\nu=0}^2 g_{\nu}(N) P_{\mu+3-\nu},
        \label{eq:boson1-2} \\       
  g_{\nu}(N)  & = & \Hb_{\nu+1} = f^2(N+\nu) F(N+\nu), \qquad \nu=0, 1, \qquad
        g_2(N) = \Hb_3,  \label{eq:boson1-3}
\end{eqnarray}
in the former case and
\begin{eqnarray}
  \Qb_{\mu} & = & c\sqrt{2} \left[f_1(N) \ap + {\rm i} f_2(N+1) a\right] P_{\mu+2}, 
        \label{eq:boson2-1}\\
  \Qbp_{\mu} & = & c\sqrt{2} \left[f_1(N+1) a P_{\mu} - {\rm i} f_2(N) \ap P_{\mu+1}
        \right], \\
  \bar{\cal H}_{\mu} & = & \sum_{\nu=0}^2 g_{\nu}(N) P_{\mu+3-\nu},
        \label{eq:boson2-3} \\       
  g_{\nu}(N)  & = & \Hb_{\nu+1} \nonumber \\ 
  & = &\case{1}{2}\left[f_1^2(N+\nu) F(N+\nu) + f_2^2(N+\nu-1) F(N+\nu-1)\right],
        \label{eq:boson2-4}
\end{eqnarray}
in the latter. Note that in (\ref{eq:boson1-3}), $g_2(N)$ remains arbitrary.\par
%
%
It is straightforward to check that the operators $\Qb_{\mu}$, $\Qbp_{\mu}$,
$\bar{\cal H}_{\mu}$ satisfy the defining relations (\ref{eq:PsSSQM-1}) --
(\ref{eq:PsSSQM-3}) of PsSSQM for any $\mu \in \{0, 1, 2\}$, any GDOA \AG (i.e., any
structure function $F(N)$), and any choice of functions $f(N)$ in (\ref{eq:Qb-1}) or
$f_1(N)$, $f_2(N)$ in (\ref{eq:Qb-2}) and (\ref{eq:Qbp-2}). Both sets of equations
(\ref{eq:boson1-1}) -- (\ref{eq:boson1-3}) and (\ref{eq:boson2-1}) --
(\ref{eq:boson2-4}) provide us with a bosonization of PsSSQM similar to that known for
SSQM~\cite{plyu} in terms of the Calogero-Vasiliev algebra~\cite{vasiliev}. It should be
stressed that such results remain valid in the case of the standard oscillator algebra ${\cal
A}(I)$.\par
%
%
\section{\boldmath Pseudosupersymmetric Quantum Mechanics in Terms of
$C_3$-Extended Oscillator Algebra Generators}

\setcounter{equation}{0}

Let us specialize the results of Secs.~5 and 6 by selecting the GDOA \Athree\ associated
with a $C_3$-extended oscillator algebra ${\cal A}^{(3)}_{\alpha_0
\alpha_1}$, where $C_3 = \Z_3$ is the cyclic group of order three~\cite{cq00, cq98}. As
generator of the latter we take the operator $T$ defined in~(\ref{eq:T}), hence $C_3 =
\{T, T^2, T^3 = I\}$.\footnote{In ${\cal A}^{(3)}_{\alpha_0 \alpha_1}$, $T$ is
considered as an operator independent of the remaining ones, so that the property $T^3 =
I$ and Eq.~(\ref{eq:T-prop}) (or alternatively Eqs.~(\ref{eq:P-prop1}) and
(\ref{eq:P-prop2})) have to be postulated in addition to (\ref{eq:GDOA}) and
(\ref{eq:G-1}) (or (\ref{eq:G-2})). The GDOA \Athree\ corresponds to the realization
(\ref{eq:T}) of $T$.}  The GDOA \Athree\ corresponds to the choice
\begin{equation}
  G(N) = I + \kappa_1 T + \kappa_2 T^2  \label{eq:G-1}
\end{equation}
in Eq.~(\ref{eq:GDOA}). Here $\kappa_1$ is some complex constant and $\kappa_2 =
\kappa_1^*$.\par
%
%
The function $G(N)$ of Eq.~(\ref{eq:G-1}) can be alternatively written in terms of the
projection operators (\ref{eq:P}) as 
\begin{equation}
  G(N) = I + \sum_{\mu=0}^2 \alpha_{\mu} P_{\mu},  \label{eq:G-2}
\end{equation}
where 
\begin{equation}
  \alpha_{\mu} = \sum_{\nu=1}^2 e^{2\pi{\rm i}\mu\nu/3} \kappa_{\nu}, \qquad
  \mu = 0, 1, 2,
\end{equation}
are some real parameters subject to the condition
\begin{equation}
  \sum_{\mu=0}^2 \alpha_{\mu} = 0.  \label{eq:alpha-cond1}
\end{equation}
The $P_{\mu}$'s can now be interpreted as the projection operators on the carrier
spaces of the three inequivalent unitary irreducible representations of the cyclic group
$C_3$.\par
%
%
{}From (\ref{eq:G-2}) and (\ref{eq:alpha-cond1}), it follows that the algebra \Athree\
depends upon two independent, real parameters $\alpha_0$, $\alpha_1$, and goes over
to the standard oscillator algebra ${\cal A}(I)$ for $\alpha_0$, $\alpha_1 \to 0$. Its
structure function $F(N)$, which is a solution of Eq.~(\ref{eq:G-F}) with $G(N)$ given
by (\ref{eq:G-2}), can be expressed as 
\begin{equation}
  F(N) = N + \sum_{\mu=0}^2 \beta_{\mu} P_{\mu}, \qquad \beta_0 \equiv 0, \qquad
  \beta_{\mu} \equiv \sum_{\nu=0}^{\mu-1} \alpha_{\nu}, \qquad \mu=1, 2.
  \label{eq:F}
\end{equation}
The existence of a bosonic Fock space representation is guaranteed by the constraints
\begin{equation}
  \alpha_0 > -1, \qquad \alpha_0 + \alpha_1 > -2  \label{eq:alpha-cond2}
\end{equation}
on the parameters, ensuring that $F(n) > 0$ for $n=1$, 2, 3,~\ldots.\par
%
%
By using the explicit form of the structure function of \Athree, given in (\ref{eq:F}), the
eigenvalues $\Eb^{(\mu)}_n$ of the bosonized pseudosupersymmetric Hamiltonian
$\bar{\cal H}_{\mu}$, defined in (\ref{eq:boson1-2}) and (\ref{eq:boson1-3}), can be
written in the form
\begin{eqnarray}
  \Eb^{(0)}_{3k} & = & 3k f^2(3k), \qquad \Eb^{(0)}_{3k+2} = \Eb^{(0)}_{3k+3},
       \qquad \Eb^{(0)}_{3k+1} \mbox{\rm \ arbitrary},  \label{eq:spectrum1-1} \\
  \Eb^{(1)}_{3k} & = & \Eb^{(1)}_{3k+1} = (3k + 1 + 2\gamma_0) f^2(3k+1), \qquad
       \Eb^{(1)}_{3k+2} \mbox{\rm \ arbitrary},  \label{eq:spectrum1-2} \\
  \Eb^{(2)}_{3k+1} & = & \Eb^{(2)}_{3k+2} = (3k + 2 + 2\gamma_2) f^2(3k+2),
       \qquad \Eb^{(2)}_{3k} \mbox{\rm \ arbitrary},  \label{eq:spectrum1-3} 
\end{eqnarray}
where
\begin{equation}
  \gamma_{\mu} \equiv \case{1}{2}(\beta_{\mu} + \beta_{\mu+1}) = \left\{
  \begin{array}{ll}
      \frac{1}{2}\alpha_0 & {\rm if\ } \mu=0\\[0.2cm]
      \alpha_0 + \frac{1}{2} \alpha_1 & {\rm if\ } \mu=1\\[0.2cm]
      \frac{1}{2} (\alpha_0 + \alpha_1) & {\rm if\ } \mu=2
  \end{array}\right..  \label{eq:gamma}
\end{equation}
Similarly, for $\bar{{\cal H}}_{\mu}$, defined in (\ref{eq:boson2-3}) and
(\ref{eq:boson2-4}), we obtain
\begin{eqnarray}
  \Eb^{(0)}_{3k} & = & \case{1}{2} [3k f_1^2(3k) + (3k-1+2\gamma_2) f_2^2(3k-1)],
       \nonumber \\
  \Eb^{(0)}_{3k+1} & = & \Eb^{(0)}_{3k+2} = \Eb^{(0)}_{3k+3}, 
       \label{eq:spectrum2-1} \\
  \Eb^{(1)}_{3k} & = & \Eb^{(1)}_{3k+1} =  \case{1}{2} [(3k+1+2\gamma_0)
       f_1^2(3k+1) + 3k f_2^2(3k)], \nonumber \\
  \Eb^{(1)}_{3k+2}  & = & \Eb^{(1)}_{3k+3}, \label{eq:spectrum2-2} \\
  \Eb^{(2)}_{3k} & = & \Eb^{(2)}_{3k+1} = \Eb^{(2)}_{3k+2} = \case{1}{2} [(3k+2
       +2\gamma_2)f_1^2(3k+2) \nonumber \\
  && \mbox{} + (3k+1+2\gamma_0) f_2^2(3k+1)]. \label{eq:spectrum2-3}
\end{eqnarray} 
\par
%
%
Such relations show that the existence of the two different types of bosonization
(\ref{eq:boson1-1}) -- (\ref{eq:boson1-3}) and (\ref{eq:boson2-1}) --
(\ref{eq:boson2-4}), as well as the arbitrariness of $f(N)$ and $f_1(N)$, $f_2(N)$ make
it virtually possible to reproduce any nonlinear pseudosupersymmetric spectra.\par
%
%
In the particular cases where $f(N)=1$ or $f_1(N) = f_2(N) = 1$, we get back the results
previously obtained for linear spectra~\cite{cq00}. Equation (\ref{eq:boson2-1}) indeed
corresponds to Eq.~(5.20) of Ref.~\cite{cq00} (for the choice $\varphi = \pi/2$ in
Eq.~(5.18) of the same reference), while Eq.~(\ref{eq:boson1-1}) can be deduced from
Eq.~(5.23) of Ref.~\cite{cq00} by interchanging the roles of $\Qb_{\mu}$ and
$\Qbp_{\mu}$ (which leaves the PsSSQM algebra invariant) and changing the label
$\mu$. The corresponding pseudosupersymmetric Hamiltonians and their eigenvalues can
be similarly related.\par
%
%
More generally, it results from Eqs.~(\ref{eq:alpha-cond2}) and (\ref{eq:gamma}) that if
$f^2(N)$ is chosen to be an increasing function of $N$ and $\Hb_3$ is a
positive-definite operator, the spectrum (\ref{eq:spectrum1-1}) has a nondegenerate
ground state at vanishing energy, whereas for the spectra (\ref{eq:spectrum1-2}) and
(\ref{eq:spectrum1-3}), the ground state at a positive energy may be nondegenerate or
twofold degenerate. Furthermore, if $f_1^2(N)$ and $f_2^2(N)$ are chosen to be 
increasing functions of $N$, the ground state of the spectrum
(\ref{eq:spectrum2-1}) is nondegenerate and its energy may have any sign, while
that of the spectra (\ref{eq:spectrum2-2}) and (\ref{eq:spectrum2-3}) is twofold or
threefold degenerate and has a positive energy.\par
%
%
\section{Summary}

In the present paper, we have constructed the complete explicit solution for the
realization of PsSSQM in terms of two superpotentials and we have established that it can
be separated into two branches corresponding to equal or unequal superpotentials,
respectively.\par
%
%
We have then proved that any order-$p$ orthosupersymmetric system has a
pseudosupersymmetry, but that the reverse is not true. We have actually given conditions
under which a pseudosupersymmetric system may be described by orthosupersymmetries
of order two. In this way, we have extended to PsSSQM some recent results valid for
PSSQM and FSSQM~\cite{agha}, thereby establishing that OSSQM is contained in the
other variants of SSQM mentioned in Sec.~1.\par
%
%
Next, we have proposed two new matrix realizations of PsSSQM in terms of GDOA
generators and we have related them to the two distinct realizations of the same in terms
of superpotentials.\par
%
%
{}Finally we have demonstrated that such matrix realizations are fully reducible and that
their irreducible components provide two distinct sets of bosonized operators realizing
PsSSQM and corresponding to nonlinear spectra. These two sets reduce to those found in
Ref.~\cite{cq00} when we choose a GDOA associated with a $C_3$-extended oscillator
algebra and restrict ourselves to linear spectra. Such results are part of a more general
study, wherein we plan to prove the full reducibility of SSQM variants and their resultant
bosonization when they are realized in terms of GDOA generators (see also
Ref.~\cite{cq02}).\par
%
%

 \end{document}